\pdfoutput=1
\documentclass[11pt]{article}
\usepackage{acl}
\usepackage{times}
\usepackage{latexsym}
\usepackage[T1]{fontenc}
\usepackage[utf8]{inputenc}
\usepackage{microtype}
\usepackage{inconsolata}
\usepackage{graphicx}
\usepackage{amsmath}
\usepackage{amsthm}
\usepackage{amssymb}
\usepackage{booktabs}
\usepackage{algorithm}
\usepackage{algpseudocode}
\usepackage{float}
\usepackage{pifont}
\usepackage{booktabs}
\usepackage{bbding}
\usepackage{makecell}
\usepackage{fancyvrb}
\usepackage{subcaption}
\usepackage{listings}
\usepackage{multirow}
\usepackage{booktabs}
\usepackage{multicol}
\usepackage[table]{xcolor}
\definecolor{lightblue}{rgb}{0.7, 0.85, 1}
\definecolor{verylightgray}{rgb}{.97,.97,.97}

\title{SCLA: Automated Smart Contract Summarization via LLMs and Control Flow Prompt}

\author{
    Xiaoqi Li$^{1}$ \quad Yingjie Mao$^{1*}$ \quad Zexin Lu$^{2}$ \quad Wenkai Li$^{1}$ \quad Zongwei Li$^{1}$\\
    $^1$Hainan University, China \\
    $^2$Hong Kong Polytechnic University, China \\
    \texttt{csxqli@ieee.org} \\ 
    \texttt{yingjiemao@hainanu.edu.cn} \\
    \texttt{zexin.lu@connect.polyu.hk} \\
    \texttt{\{cswkli, lizw1017\}@hainanu.edu.cn} \\
    \textsuperscript{*}Corresponding author
}

\begin{document}
\maketitle
\begin{abstract}
Smart contract code summarization is crucial for efficient maintenance and vulnerability mitigation. While many studies use Large Language Models (LLMs) for summarization, their performance still falls short compared to fine-tuned models like CodeT5+ and CodeBERT. Some approaches combine LLMs with data flow analysis but fail to fully capture the hierarchy and control structures of the code, leading to information loss and degraded summarization quality. We propose SCLA, an LLMs-based method that enhances summarization by integrating a Control Flow Graph (CFG) and semantic facts from the code’s control flow into a semantically enriched prompt. SCLA uses a control flow extraction algorithm to derive control flows from semantic nodes in the Abstract Syntax Tree (AST) and constructs the corresponding CFG. Code semantic facts refer to both explicit and implicit information within the AST that is relevant to smart contracts. This method enables LLMs better to capture the structural and contextual dependencies of the code. We validate the effectiveness of SCLA through comprehensive experiments on a dataset of 40,000 real-world smart contracts. The experiment shows that SCLA significantly improves summarization quality, outperforming the SOTA baselines with improvements of 26.7\%, 23.2\%, 16.7\%, and 14.7\% in BLEU-4, METEOR, ROUGE-L, and BLEURT scores, respectively.\par
\end{abstract}
\section{Introduction}
Smart contracts~\cite{smart_contract} are self-executing programs on Ethereum, and the blockchain’s immutability complicates vulnerability maintenance~\cite{Authros}. Solidity, designed specifically for smart contract development, compiles code into bytecode and ABI for execution on the Ethereum Virtual Machine (EVM). Unlike general-purpose languages like Java and Python, Solidity emphasizes security with strict type safety and single-threaded execution. Analyzing Solidity code requires examining syntax, semantics, and state management. Even minor vulnerabilities can result in financial losses~\cite{Vulnerabilities}, making smart contract code summarization essential for improving efficiency and reducing security risks. Smart contract summarization has received less attention than Java and Python, with traditional methods relying on deep learning and fine-tuning. Yang et al.~\cite{MMTran} proposed MMTrans, integrating deep learning with structure-based traversal (SBT) and Abstract Syntax Tree (AST) graphs for code summarization. Lei et al.~\cite{FMCF} introduced FMCF, a Transformer-based method that fuses multi-scale features to preserve both semantic and syntactic information. Zhao et al.~\cite{comment_generation} proposed SCCLLM, combining context learning with information retrieval to improve summarization.\par
However, these approaches pose security risks. Fine-tuning is limited by the quality and size of training data, which may lack emerging vulnerabilities. Additionally, fine-tuned models may suffer from knowledge forgetting~\cite{forget_kon}, reducing adaptability. In contrast, LLMs-based approaches leverage extensive pre-trained datasets to identify vulnerabilities and enhance security. However, prior methods mainly use isolated code snippets, failing to capture the full semantic context. Ahmed et al.~\cite{Automatic_Semantic} demonstrated that LLMs struggle with implicit semantics, leading to the loss of critical information. These methods also neglect control structures, hindering accurate summarization and security.\par
\begin{figure*}[ht]
    \centering
    \includegraphics[width=0.99\linewidth]{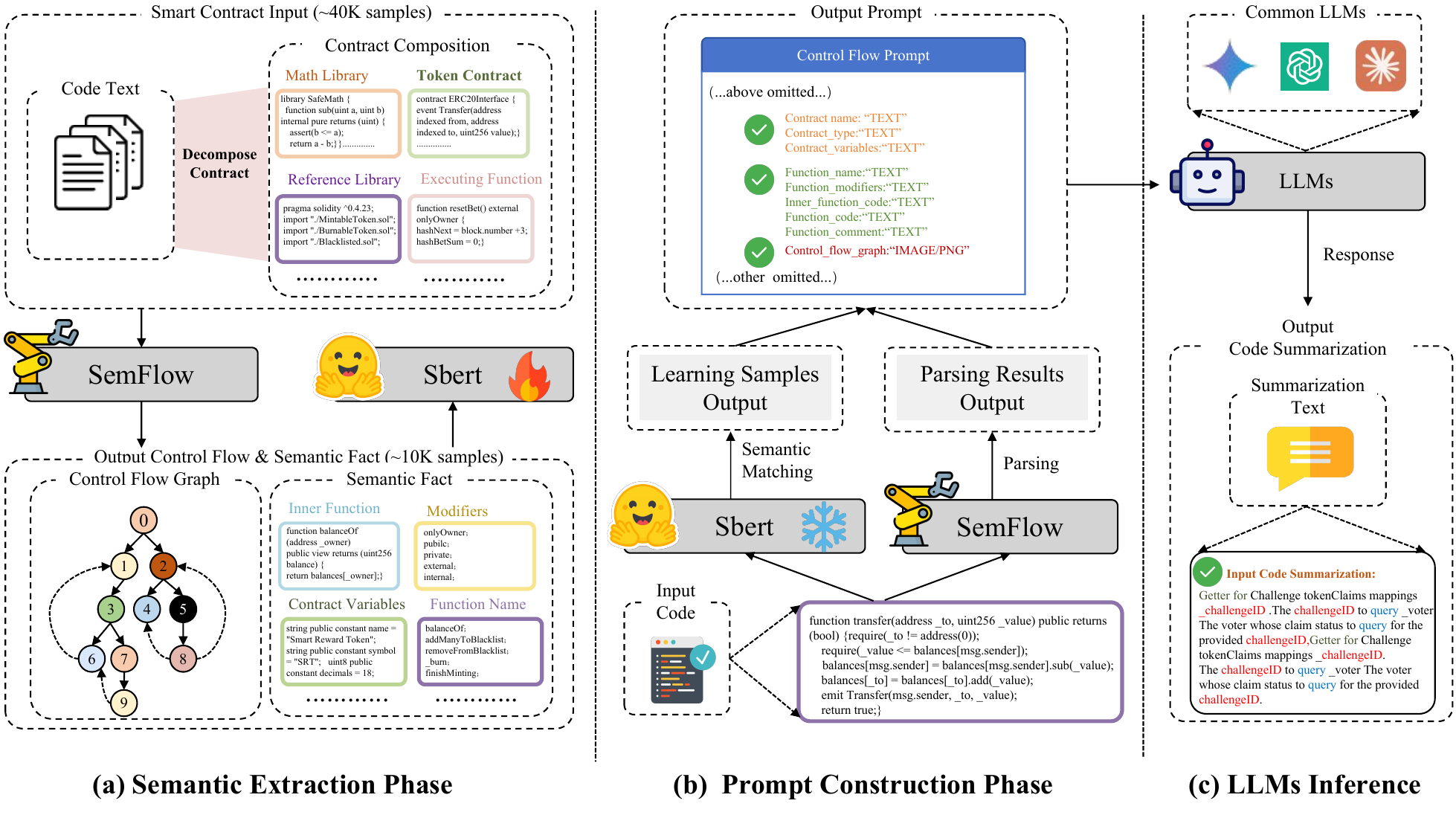}
    \caption{Overview of our proposed framework, SCLA, powered by Google's Gemini-1.5-Pro, performs automated generation of smart contract code summarization. SCLA extracts control flow semantic facts from smart contract code and uses Gemini-1.5-Pro to generate code summarization from control flow semantic facts.}
    \label{framework}
\end{figure*}
To address the limitations of existing methods and generate more secure smart contract code summarization, we propose SCLA (Smart Contract summarization with LLMs and Semantic Augmentation). SCLA integrates LLMs with control flow analysis to enhance scalability and summarization quality. Incorporating control flow-based semantic information and control flow graphs improves security and semantic accuracy in LLMs-generated summarization. The approach extracts this information via semantic analysis and generates control flow graphs. Using few-shot prompting and a fine-tuned Sentence-Transformer\cite{sbert}, SCLA identifies semantically similar examples for task-specific prompts. \textbf{SemFlow}, a core component, extracts control flow graphs and semantic details, such as function call graphs and variables. To prevent overwhelming LLMs, non-control flow information is presented separately, while control flow graphs are provided in a tagged PNG format. Our experiments on 14,789 method-comment pairs from a GitHub repository of 40,000 smart contracts show that control flow graphs improve LLMs' performance.\par
Our main contributions can be summarized as follows:
\begin{itemize}
    \item We propose SCLA, the first framework that integrates LLMs with smart contract code summarization using control flow prompts. It extracts control flow graphs and associated semantic information from the AST, enhancing the LLM's understanding of code structure.
    \item We conduct extensive experiments on a dataset with 14,795 method-comment pairs, using BLEU-4, METEOR, ROUGE-L, and BLEURT as evaluation metrics. We perform a comparative analysis with state-of-the-art approaches, achieving a 37.53 BLEU-4 score, 52.54 METEOR score, 56.97 ROUGE-L score, and 63.4 BLEURT score.
    \item We thoroughly evaluated the generalizability of SCLA through extensive experiments on Java and Python datasets, offering valuable insights for future research on control flow-based prompts in other code domains.
    \item We have uploaded the related code and experimental data to Figshare, with plans to open-source them after the paper's publication.
\end{itemize}
\section{Related Work}
\subsection*{Smart Contract Summarization}
Deep learning models have made significant advances in smart contract code summarization. Yang et al.~\cite{MMTran} proposed MMTrans, which extracts SBT sequences and AST-based graphs to capture global and local semantics using dual encoders and a joint decoder. Transformer models like CodeT5~\cite{codet5} and CodeBERT~\cite{code_best} also enhance summarization quality but require extensive fine-tuning and large datasets. LLMs, such as GPT-4o and Gemini-1.5-Pro, excel in few-shot or zero-shot summarization tasks, bypassing fine-tuning. Previous studies~\cite{Few-shot,Automatic_Semantic} highlight the benefits of few-shot learning. However, LLMs often produce suboptimal summarization, lacking conciseness and functional generalization. Ahmed et al.~\cite{Automatic_Semantic} proposed ASAP, incorporating data flow and GitHub context. Still, it fails to capture function call relationships and control flow, suggesting the need for improved semantic facts and control flow integration for better summarization.\par
\section{METHODOLOGY}
\begin{figure}[ht]
	\centering
        \includegraphics[width=0.99\linewidth]{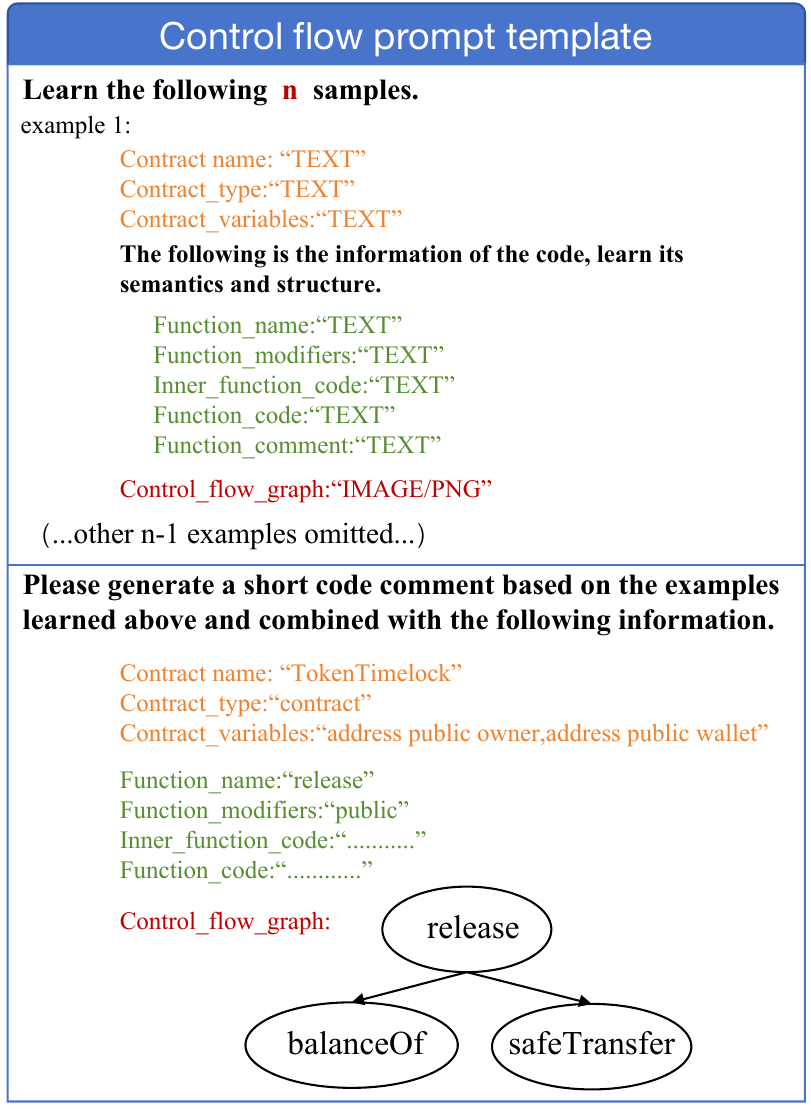}
	\caption{An Example of Control Flow Prompt.}
	\label{example}
\end{figure}
\label{methodology}
\subsection{Control Flow Prompt}
In this section, we discuss the control flow prompt and the corresponding semantic facts utilized by SCLA, as illustrated in Figure \ref{example}. These semantic facts are carefully integrated into the prompts to enhance the LLMs' ability to generate more accurate, relevant, and comprehensive summarizations.\par
\textbf{\textit{Control Flow Graph \& Inner Function}}.
We define the set of inner functions as those invoked within the target function, with each element referred to as an inward function. The function call graph captures the precise sequence of function calls, representing the control flow of the target code. This graph is used as control flow input for the LLMs, along with the set of inner functions, to provide valuable additional context about invoked functions. This approach mitigates misinterpretation based solely on function names, significantly enhancing semantic inference. Moreover, the function call graph helps the LLMs accurately determine the sequence and depth of function calls, thereby aiding in the understanding of complex functions and their interdependencies.\par

\textbf{\textit{Identifiers}}. Previous studies highlight that identifiers play a critical role in helping language models retrieve valuable information for code summarization~\cite{role_of_identifiers}. Identifiers, including modifiers, local variables, and function names, offer essential context about the code's operations. By understanding an identifier’s role, the language model can better interpret the code. In our approach, we use a tree-sitter to traverse the AST of the function, collect identifiers and their roles, and incorporate them into the prompt.\par

\textbf{\textit{Contract Name \& Global Member Variables}}. Incorporating domain-specific information into prompts greatly enhances LLMs' overall performance and effectiveness, particularly in specialized tasks such as smart contract analysis. For instance, smart contract names~\cite{Solana} often reflect their functional roles or token names~\cite{smart_contract_name}, providing valuable contextual information for the LLMs. Additionally, global member variables, such as contract addresses and account balances, assist LLMs in more effectively understanding contract functions and their interrelations. This significantly reduces the need for LLMs to infer complex operations from variable names, leading to more precise descriptions and significantly improved summarization accuracy.\par

\subsection{Semantic-based Retrieval}
In this paper, we use the \textbf{Sentence-Transformer} (Sbert)\cite{sbert} model to semantically match the code samples in the repository that are most similar to the target code snippet, which is then used as few-shot learning examples in the prompt. First, we divide the samples in the code repository into three subsets: training set, test set, and validation set (as shown in Table \ref{dataset_table}), and fine-tune the SBERT model using the training set. We begin by vectorizing the given sentences \( S_1 \) and \( S_2 \), as described by the following formula:
\begin{equation}
\begin{aligned}
\mathbf{v}_1 &= \text{Pooling}(BERT(S_1)) \\
\mathbf{v}_2 &= \text{Pooling}(BERT(S_2))
\end{aligned}
\end{equation}
\par
Sbert is trained using contrastive learning or triplet loss, optimizing sentence embeddings such that similar sentences are closer in vector space and dissimilar sentences are farther apart. Given a positive pair \( (S_1, S_2) \) and a negative pair \( (S_1, S_3) \), the model optimizes the following loss function:
\begin{equation}
\begin{aligned}
\mathcal{L} = \max \left( 0, 
\text{cosine\_similarity}(\mathbf{v}_1, \mathbf{v}_2) \right. \\
\left. - \text{cosine\_similarity}(\mathbf{v}_1, \mathbf{v}_3) + \Delta \right)
\end{aligned}
\end{equation}
where \( \Delta \) is a margin hyperparameter that controls the minimum desired similarity difference, and \( \mathbf{v}_1 \) and \( \mathbf{v}_2 \) are the vector representations of sentences \( S_1 \) and \( S_2 \).\par
Finally, we compute the cosine similarity between the target code vector and the repository code vectors to identify the most semantically similar samples. The formula is as follows:
\begin{equation}
\text{cosine\_similarity}(\mathbf{v}_1, \mathbf{v}_2) = \frac{\mathbf{v}_1 \cdot \mathbf{v}_2}{\|\mathbf{v}_1\| \|\mathbf{v}_2\|}
\end{equation}

For each target sample, we rank the repository samples by cosine similarity in descending order. The top \(k\) matches, as specified by the parameter \(\text{number\_top\_matches}\), are selected and stored in a result dictionary, which contains the matched code snippets and their similarity scores. If a file path is provided, the results are serialized and saved in JSON format for further analysis or review.\par

\subsection{SCLA Framework}
Figure \ref{framework} illustrates the overall framework of SCLA. We outline the three stages of the SCLA process for generating smart contract code summarization.\par
\textbf{Semantic Extraction:} We use local extraction methods to segment .sol files, avoiding parsing errors. Regular expressions extract code and comments, which are then passed to \textbf{SemFlow} for semantic extraction. The function call graph and semantic facts are stored in a repository, indexed by contract file path, and named UUIDs. \par
\textbf{Prompt Construction:} SCLA uses few-shot learning to enhance LLMs' code summarization performance. Sentence-Transformer~\cite{sbert} retrieves the top \(k\) semantically similar code samples. The extracted semantic information, including function call graphs, function arguments, function modifiers, and contract metadata, is integrated into the prompt. \par
\textbf{LLMs Inference:} The semantically enhanced prompt, including the function call graph, is input into the LLMs interface to improve understanding of the function call sequence, resulting in higher-quality code summarization.\par

\begin{algorithm}[h]
\caption{Source Data to Function Call Tree}
\label{ASTtograph}
\scriptsize
\begin{algorithmic}[1]
\State \textbf{Input:} Source code $f$ to be parsed by Solparser; initialized empty dictionary $T$
\State \textbf{Output:} Function call tree $T$
\State $AST \gets \text{Solparser.parser}(f)$
\State $T \gets \{\}$
\For{each $c$ in $AST$}
    \For{each $g$ in $c$}
        \For{each $x$ in $g.\text{calls}$}
            \State $n \gets x.\text{name}$
            \If{$n \notin T[c][g]$}
                \State $T[c][g][n] \gets \{c: c, \text{count}: 1\}$
            \Else
                \State $T[c][g][n].\text{count} \gets T[c][g][n].\text{count} + 1$
            \EndIf
        \EndFor
    \EndFor
\EndFor
\For{each $c$, $g$ in $T$}
    \State \textbf{CreateCallTree}($c$, $g$, $T[c][g]$, $T$)
\EndFor
\State \Return $T$

\Function{CreateCallTree}{p, k, n, T}
    \For{each $m$ in $n.\text{keys}$}
        \State $o \gets n[m]$
        \If{$m \notin T[p][k].\text{keys}$}
            \State $T[p][k][m] \gets T[o.\text{c}][m]$
            \State \textbf{CreateCallTree}($k$, $m$, $T[o.\text{c}][m]$, $T$)
        \EndIf
    \EndFor
\EndFunction
\end{algorithmic}
\end{algorithm}
\subsection{Control Flow Extraction}
We use \textbf{SemFlow}, a component integrated with a control flow extraction algorithm, to extract function call graphs from the AST as control flow input in the prompt." The algorithm in \ref{ASTtograph} demonstrates the entire extraction process. It first uses an AST parsing tool to parse the input code into an AST. The AST is traversed in a depth-first manner to remove irrelevant nodes, such as imports. Function nodes with calls are marked in the "FunctionCall" field, allowing the construction of a reference tree (lines 5-20 of Algorithm \ref{ASTtograph}). The depth of the reference tree ranges from 2 to 3 layers, depending on the presence of function calls. When a third-level call points to a second or third-level node, the reference tree is transformed into a complete call tree by grafting branch nodes (lines 21-29 of Algorithm \ref{ASTtograph}). The call tree is then visualized using Graphviz and saved to the code sample repository.\par
\begin{table}[htp!]
    \centering
    \setlength{\abovecaptionskip}{0.cm}
    \resizebox{0.99\linewidth}{!}{
    \begin{tabular}{c|c|c|c}
    \toprule
        \textbf{Type} &\textbf{Train} & \textbf{Validation} &\textbf{Test} \\  \hline
        Number &11032 &2758 &1000\\
        Avg. tokens in codes &42.44 &42.08  &41.95 \\
        Avg. tokens in comments &26.34 &26.16  &26.66 \\
        \hline
    \end{tabular}}
    \caption{Statistics of Experimental Dataset.}
    \label{dataset_table}
\end{table}
\begin{table*}[h]
    \renewcommand{\arraystretch}{1.2}
    \centering
    \setlength{\abovecaptionskip}{0.cm}
    \resizebox{0.99\linewidth}{!}{
    \normalsize
    \begin{tabular}{c|c|c|c|c|c|c|c|c|c|c|c}
    \toprule
        \multirow{2}{*}{\textbf{Model}} & \multirow{2}{*}{\textbf{\# of sample}} 
        & \multicolumn{3}{c|}{\textbf{BLEU-4}} & \multicolumn{3}{c|}{\textbf{METEOR}} & \multicolumn{3}{c|}{\textbf{ROUGE-L}} & \multirow{2}{*}{\textbf{p-value}}  \\ \cline{3-11}
        && \textbf{Zero-Shot} & \textbf{+CFG +IF} & \textbf{Gain(\%)} 
        & \textbf{Zero-Shot} & \textbf{+CFG +IF} & \textbf{Gain(\%)} 
        & \textbf{Zero-Shot} & \textbf{+CFG +IF} & \textbf{Gain(\%)} \\\hline
        Llama-3.2-1b-preview & 11032 & 3.03 & \textbf{5.43} & +79.21\%  & 19.58 & \textbf{23.97} & +22.42 & 18.88 &\textbf{23.49} & +24.42 & <0.01 \\
        GPT-4o & 11032 & 5.34 & \textbf{7.45} & +39.51\%  & 22.32 & \textbf{26.62} & +19.27 & 25.32 & \textbf{32.62} & +28.83 & <0.01 \\
        Gemini-1.0-Pro-Vision & 11032 & 3.01 & \textbf{5.32} & +76.74\%  & 16.89 & \textbf{20.73} & +22.73 & 18.46 & \textbf{20.31} & +10.02 & <0.01 \\
        Gemini-1.5-Pro & 11032 & 3.21 & \textbf{5.87} & +82.87\%  & 19.89 & \textbf{25.61} & +28.76 & 23.95 & \textbf{27.42} & +14.49 & <0.01 \\
        Claude-3.5-sonnet & 11032 & 3.31 & \textbf{5.32} & +60.73\%  & 23.42 & \textbf{28.62} & +22.20  & 25.82 & \textbf{30.12} & +16.65 & <0.01 \\
    \bottomrule
    \end{tabular}}
    \caption{Performance of different LLMs on smart contract code summarization, measured using BLEU-4, METEOR, ROUGE-L. p-values are calculated applying a one-sided pairwise Wilcoxon signed-rank test and B-H corrected.}
    \label{validity_experiment}
\end{table*}
\begin{figure*}[h]
    \centering
    \begin{subfigure}{0.3\textwidth}
        \centering
        \includegraphics[width=\textwidth]{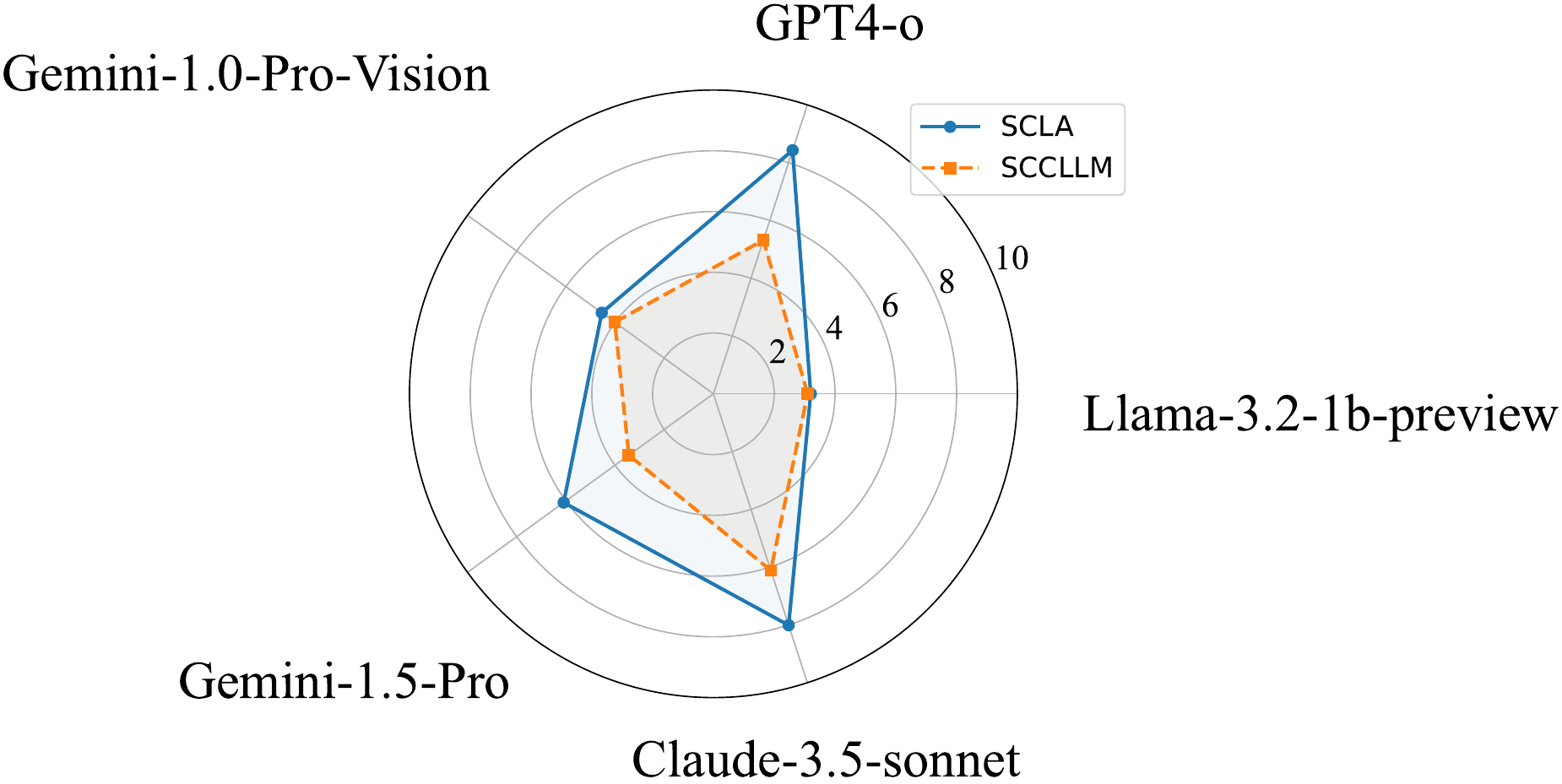}
        \caption{Comparison Results on BELU-4}
    \end{subfigure}
    \begin{subfigure}{0.3\textwidth}
        \centering
        \includegraphics[width=\textwidth]{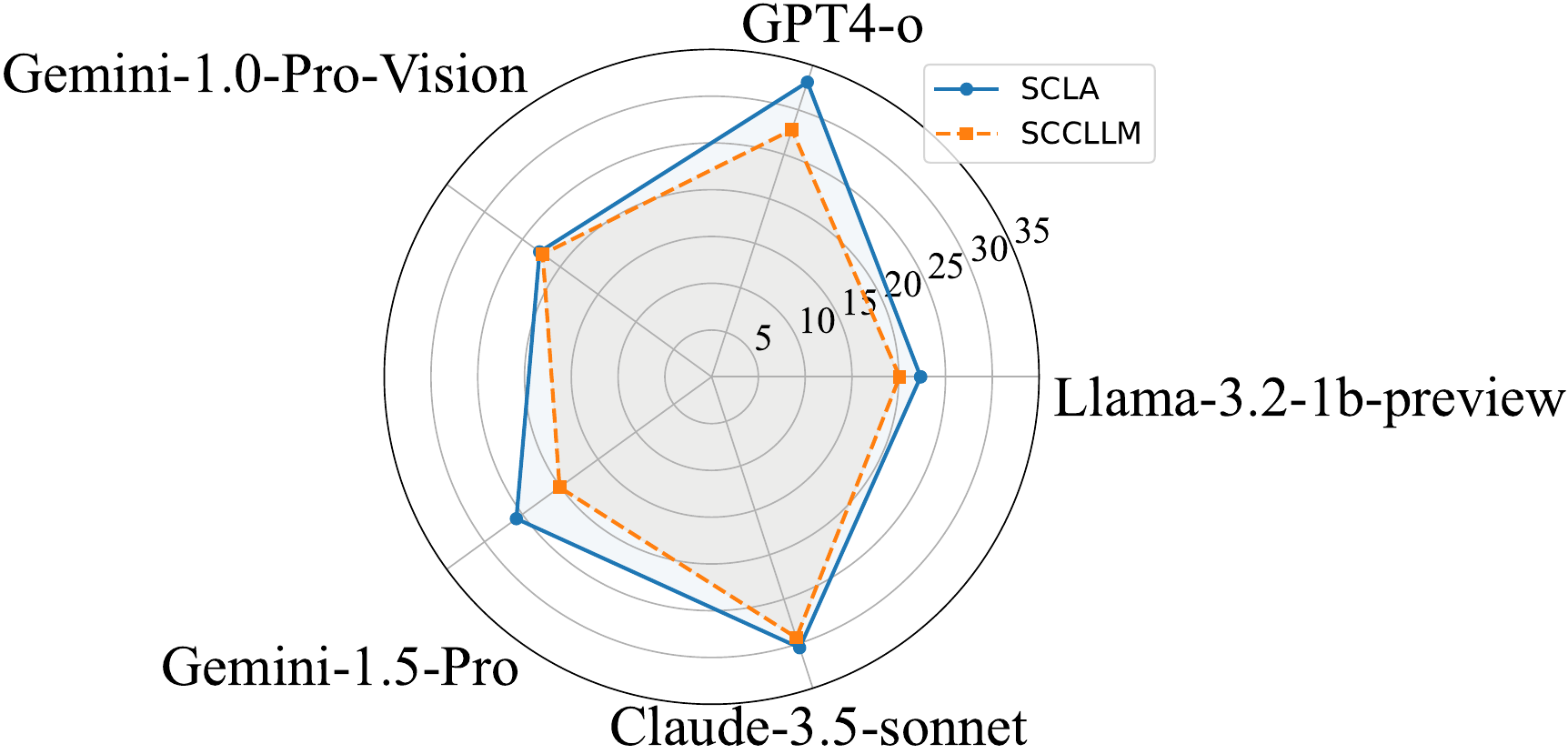}
        \caption{Comparison Results on METEOR}
    \end{subfigure}
    \begin{subfigure}{0.3\textwidth}
        \centering
        \includegraphics[width=\textwidth]{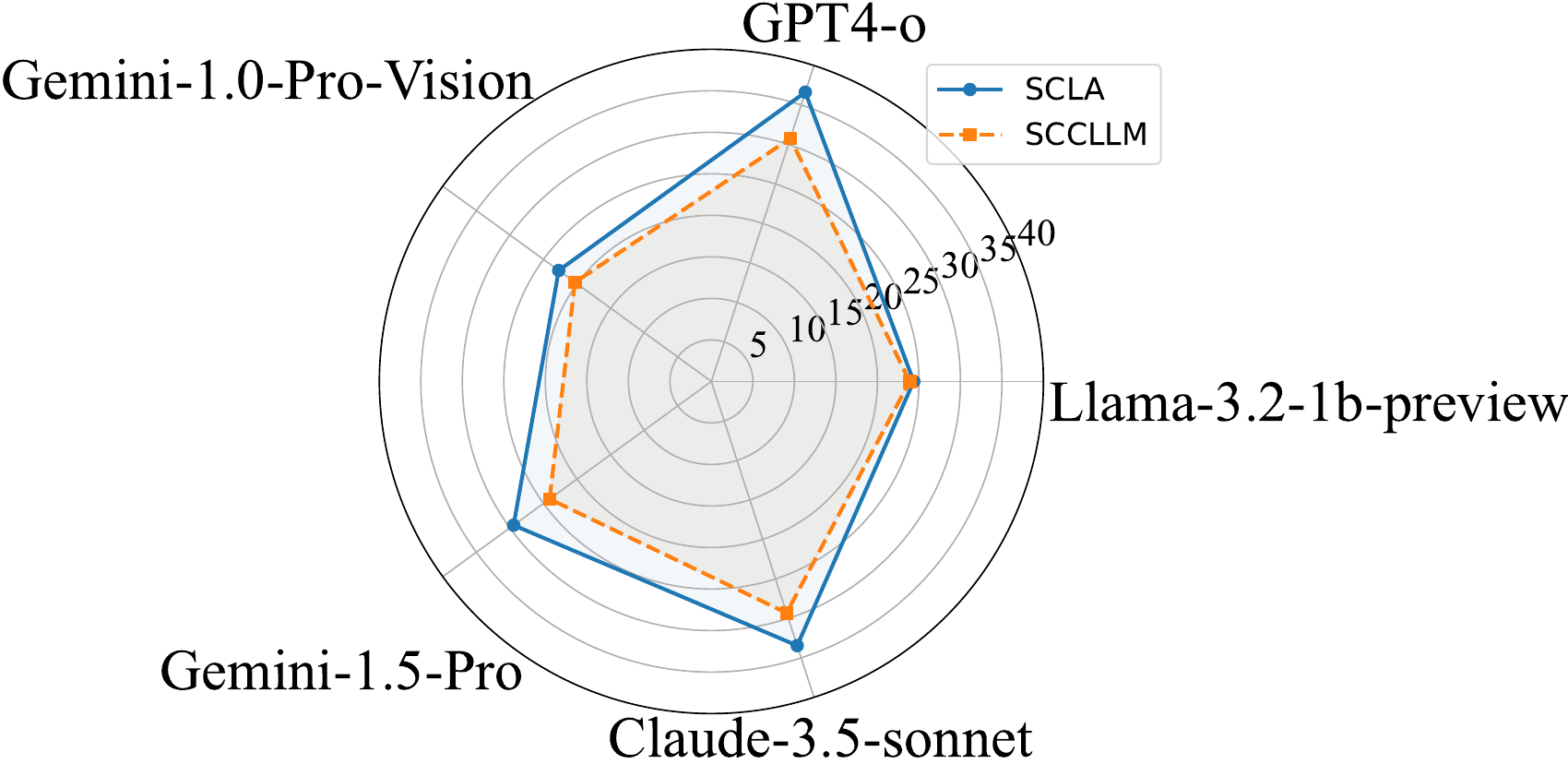}
        \caption{Comparison Results on ROUGE-L}
    \end{subfigure}
    \caption{The Comparison of BLEU, METEOR, and ROUGE-L Scores on Our Test Set Under Five Different LLMs, Using the SCCLLM  and the Proposed SCLA for Zero-Shot Summarization Tasks.}
   \label{rader_graph}
\end{figure*}
\begin{table*}[h]
    \centering
    \setlength{\abovecaptionskip}{0.cm}
    \resizebox{0.99\linewidth}{!}{
    \begin{tabular}{c|c|c|c|c|c|c|c}
    \toprule
        \textbf{Approach} &\textbf{\#of training sample} &\textbf{\#of test sample} &\textbf{BLEU-4} & \textbf{METEOR}  & \textbf{ROUGE-L} &\textbf{BLEURT} &\textbf{p-values}\\  \hline
        CodeT5+ &11032 &1000 &28.95  &45.62 &49.77 &57.79 &/ \\ 
        CodeT5 &11032 &1000&27.24 &43.31 &49.03 &52.61 &/\\
        CodeBERT &11032 &1000&26.31 &39.57 &44.52 &52.74 &/\\
        MMTran &11032 &1000 &22.12  &38.92 &40.12 &54.73 &/ \\ 
        FMCF &11032 &1000&29.98 &36.67 &51.21 &51.73 &/\\
        SCCLLM (One-Shot) &/ &1000 &19.45 &20.12 &19.12 &36.56 &<0.01 \\ 
        SCCLLM (Three-Shot) &/ &1000 &29.73 &35.33 &49.44 &50.91 &<0.01\\ 
        SCCLLM (Five-Shot) &/ &1000 &31.73 &48.12 &60.44 &58.74 &<0.01\\ \hline
        SCLA (Zero-Shot) &/ &1000 & 6.09 &25.80 &29.45 &46.63 &<0.01\\
        SCLA (One-Shot) &/ &1000 & 25.46 &42.78 &47.55 &57.07 &<0.01\\
        \rowcolor{lightblue} \textbf{SCLA (Three-Shot)} &/ &1000 & \textbf{35.15} &\textbf{51.80} &\textbf{55.89}  & \textbf{63.11} &<0.01\\
        \textbf{SCLA (Five-Shot)} &/ &1000 & \textbf{37.53} &\textbf{52.54} &\textbf{56.97}  &\textbf{63.44} &<0.01\\\hline
    \end{tabular}}
    \caption{The impact of different few-shot learning quantities on SCLA performance with Gemini-1.5-Pro. p-values are calculated applying a one-sided pairwise Wilcoxon signed-rank test and B-H corrected.}
    \label{sample_learning}
\end{table*}

\begin{table}[h]
    \renewcommand{\arraystretch}{1.1}
    \centering
    \setlength{\abovecaptionskip}{0.cm}
    \resizebox{0.99\linewidth}{!}{
    \begin{tabular}{ccccc}
    \toprule
        \textbf{Type} &\textbf{Zero-Shot} & \textbf{One-Shot} &\textbf{Three-Shot} &\textbf{Five-Shot}\\  \hline
        Avg. tokens in prompt &561.4 &1154.8 &2242.5 &3330.3
        \\\hline
    \end{tabular}}
    \caption{Number of Tokens Consumed with Different Numbers of Learning Sample for SCLA.}
    \label{token_length}
\end{table}
\section{EXPERIMENT}
\label{experiment}
In the empirical study, we conducted comparison, ablation, and generalization experiments. First, we used \textbf{SemFlow} to process the raw data and generate semantic facts, data flow graphs, and the semantic sample library. The code snippets were then input into SCLA for summarization and evaluation. In the comparison experiment, we varied the number of few-shot learning samples and compared the evaluation scores with baseline methods. Ablation experiments assessed the contribution of different semantic components, while generalization experiments extended SCLA to Java and Python code summarization tasks. The results and expert evaluations validate the effectiveness of SCLA in generating smart contract code summarizations.\par
\subsection{Experiment Settings}
\label{Experiment_Settings}
All our experiments are performed on a computer equipped with an Nvidia GeForce RTX 4070Ti GPU (12GB graphic memory), Gen Intel (R) Core (TM) i9-13900K, running Ubuntu 22.04 LTS.\par
\subsection{DataSet}
The raw data for our study, provided by Liu et al.\cite{dataset}, consists of 40,000 Solidity smart contracts from Etherscan.io\footnote{https://etherscan.io/}, authored by professional developers and deployed on Ethereum. Building on Yang et al.'s work\cite{MMTran}, we developed an extraction method using AST location data to segment the code into distinct domains and extract functions with their documentation via regular expressions. After filtering low-quality summarizations, we removed poor entries, resulting in 14,790 method-comment pairs. The dataset is divided into 11,032 training, 2,758 validation, and 1,000 test sets, with average token counts reported. Notably, unlike baselines, our approach, SCLA, does not require a validation set.\par
\noindent
\subsection{Baseline} 
We compare our proposed SCLA with six state-of-the-art methods, including general code summarization models such as \textbf{CodeT5}~\cite{codet5}, \textbf{CodeT5+}~\cite{codet5+}, and \textbf{CodeBERT}~\cite{code_best}, deep learning-based smart contract code summarization methods \textbf{MMTran}~\cite{MMTran} and \textbf{FMCF}~\cite{FMCF}, and smart contract-specific code summarization methods based on the latest LLMs, such as \textbf{SCCLLM}~\cite{comment_generation}.\par
\noindent
\subsection{Performance Metrics}  
To evaluate SCLA performance against baselines, we adopted various automatic performance metrics, including \textbf{BLEU-4}~\cite{bleu}, \textbf{METEOR}~\cite{meteor}, \textbf{ROUGE-L}~\cite{rouge}, and \textbf{BLEURT}~\cite{BLEURT}. These metrics effectively assess the similarity between the automatically generated smart contract summarization and the real human-generated summarization. BLEURT, in particular, calculates similarity based on sentence semantics by using a pre-trained BERT model, providing a more accurate reflection of semantic meaning.

\subsection{Main Results}
We conducted a comprehensive evaluation of the LLMs-driven SCLA framework in two different experimental setups. The framework demonstrated significant performance improvements in smart contract code summarization tasks in both zero-shot and few-shot settings. These findings provide valuable insights and contributions to the research community. The specific results are as follows:\par
\textbf{Zero-shot Results}. To evaluate the impact of function call graphs and internal functions on LLMs-generated code summarization, we conducted experiments using GPT-4o, Gemini-1.5-Pro, and Claude-3.5-Sonnet under zero-shot conditions. The experiment had two phases: first, the target code was embedded into the prompt and evaluated with standard metrics; second, the prompt was enhanced with internal functions and control flow graphs, followed by re-evaluation. Table \ref{validity_experiment} shows that incorporating internal functions and call graphs improved summarization. GPT-4o improved by 39.51\%, 19.27\%, and 28.83\%; Gemini-1.5-Pro by 82.87\%, 28.76\%, and 14.49\%; and Claude-3.5-Sonnet by 60.73\%, 22.20\%, and 16.65\%. However, Gemini-1.5-Pro underperformed compared to GPT-4o and Claude-3.5-Sonnet. \textbf{These results validate our hypothesis that control flow graphs enhance smart contract summarization}. To further validate the control flow prompt's effectiveness, we compared SCLA with SCCLLM using five multimodal models on the test set. Results in Figure \ref{rader_graph} show SCLA outperforming SCCLLM in BLEU, METEOR, and ROUGE-L scores. \textbf{This demonstrates that SCLA with the control flow prompt outperforms SCCLLM, confirming the effectiveness of control flow prompts}.\par

\begin{table*}[htp]
    \centering
    \setlength{\abovecaptionskip}{0.cm}
    \resizebox{0.99\linewidth}{!}{
    \normalsize
    \begin{tabular}{c|c|c|c|c|c|c|c|c|c}
    \hline
        \multirow{2}{*}{\textbf{Language}} & \multirow{2}{*}{\textbf{Model}} & \multirow{2}{*}{\textbf{\#of Test Sample}} 
        & \multicolumn{3}{c|}{\textbf{BLEU-4}} 
        & \multicolumn{3}{c|}{\textbf{BLEURT}} & \multirow{2}{*}{\textbf{p-values}}\\
    \cline{4-9} 
        &&& \textbf{SCCLLM} & \textbf{SCLA} & \textbf{\makecell[c]{Gain (\%)}}  
        & \textbf{SCCLLM} & \textbf{SCLA} & \textbf{\makecell[c]{Gain (\%)}}\\  
    \hline
        \multirow{3}{*}{\textbf{Java}} 
        & GPT-4o & 1000 & 28.59 & \textbf{38.43} & +34.42 & 50.34 & \textbf{68.89} & +36.85 & <0.01  \\
        & Gemini-1.5-Pro & 1000 & 23.22 & \textbf{31.43} & +35.36 & 56.33 & \textbf{63.67} & +13.03 & <0.01\\
        & Claude-3.5-sonnet & 1000 & 31.05 & \textbf{39.13} & +26.02 & 58.89 & \textbf{70.90} & +20.40 & <0.01\\
    \hline
        \multirow{3}{*}{\textbf{Python}} 
        & GPT-4o & 1000 & 22.78 & \textbf{29.56} & +29.76 & 55.90 & \textbf{64.23} & +14.90 & <0.01 \\
        & Gemini-1.5-Pro & 1000 & 20.15 & \textbf{26.06} & +29.33 & 51.78 & \textbf{61.03} & +17.86 & <0.01\\
        & Claude-3.5-sonnet & 1000 & 25.45 & \textbf{33.77} & +32.69 & 58.21 & \textbf{73.56} & +26.37 & <0.01\\
    \hline
        \rowcolor{lightblue} \textbf{Overall} & / & / & 25.21 & \textbf{33.06} & \textbf{+31.14} & 55.24 & \textbf{67.05} & \textbf{+21.38} & <0.01 \\
    \hline
    \end{tabular}}
    \label{slc_vs_sccllm}
    \caption{The performance of SCLA and SCCLLM on the Java and Python tasks, driven by three different LLMs, was evaluated using BLEU-4 and BLEURT as metrics. To assess the statistical significance of the results, p-values were calculated using a one-sided pairwise Wilcoxon signed-rank test, with Benjamini-Hochberg (B-H) correction applied for multiple comparisons.}
\end{table*}
\begin{table*}[htp!]
    \renewcommand{\arraystretch}{1.2}
    \centering
    \setlength{\abovecaptionskip}{0.cm}
    \resizebox{0.99\linewidth}{!}{
    \normalsize
    \begin{tabular}{c|c|c|c|c|c|c|c|c}
    \toprule
        \multirow{2}{*}{\textbf{Approach}} & \multirow{2}{*}{\textbf{\#of Training Sample}} & \multirow{2}{*}{\textbf{\#of Test Sample}} 
        & \multicolumn{3}{c|}{\textbf{Java}} & \multicolumn{3}{c}{\textbf{Python}} \\
        \cline{4-9} 
        &&& \textbf{BLEU-4} & \textbf{METEOR}  & \textbf{ROUGE-L}
        & \textbf{BLEU-4} & \textbf{METEOR}  & \textbf{ROUGE-L} \\\hline
        CodeBERT & 8000 & 1000 & 19.91 & 25.11 & 34.34 & 20.56 & 33.37 & 33.19 \\
        CodeT5 & 8000 & 1000 & 22.45 & 28.98 & 41.98 & 28.82 & 37.98 & 39.52 \\
        CodeT5+ & 8000 & 1000 & 28.82 & 39.79 & 49.31 & 34.67 & 46.98 & 47.34 \\
        \rowcolor{lightblue} \textbf{SCLA} & / & 1000 & \textbf{34.34} & \textbf{50.66} & \textbf{60.71} & \textbf{37.34} & \textbf{52.61} & \textbf{57.49} \\
    \hline
    \end{tabular}}
    \caption{The performance of our proposed method and the baseline model was evaluated on Java and Python datasets.}
    \label{java_dataset}
\end{table*}
\begin{table}[htp!]
    \centering
    \setlength{\abovecaptionskip}{0.cm}
    \resizebox{0.99\linewidth}{!}{
    \begin{tabular}{cccccc}
    \toprule
       \textbf{Approach}&\textbf{Prompt Component} &\textbf{BLEU-4} & \textbf{METEOR}  & \textbf{ROUGE-L} &\textbf{BLEURT} \\  \hline
        \multirow{4}{*}{\textbf{SCLA}}&ALL &\textbf{6.09} &25.80 &\textbf{29.45} &\textbf{46.63}\\
        &-CFG &5.21 &27.77 &27.94 &45.90 \\
        &-IF &4.42 &25.43 &26.23 &44.56 \\
        &-Id\&MGV &5.62 &25.47 &29.01 &46.32 \\
        &-ALL &0.85 &20.28 &18.53 &43.53 \\\hline
    \end{tabular}}
    \caption{Ablation study. Effect of Semantic Augmentation on Gemini-1.5-Pro Generated Summarization. CFG is Control Flow Graph, IF is Inner Function, Id\&MGV is Identifiers\&Global Member Variables.}
    \label{test}
\end{table}

\begin{table*}[h]
    \renewcommand{\arraystretch}{1.0}
    \centering
    \small
    \setlength{\abovecaptionskip}{0.cm}
    \fontsize{24}{32}\selectfont
    \resizebox{0.99\linewidth}{!}{
    \begin{tabular}{ccccc}
    \toprule
    \multicolumn{5}{c}{\textbf{Example}} \\\hline
& \multicolumn{3}{l}{
\begin{tabular}{@{}l@{}} 
\quad\quad\quad\quad\quad \textbf{\texttt{\#Input Function Code}} \\
\quad\quad\quad\quad\quad\quad \textbf{\texttt{function transferDataOwnership (address \_addr) onlyOwner public \{}} \\
\quad\quad\quad\quad\quad\quad\quad\quad\quad\quad \textbf{\texttt{data.transferOwnership(\_addr);}} \\
\quad\quad\quad\quad\quad \textbf{\texttt{\}}} \\
\quad\quad\quad\quad\quad\quad \textbf{\texttt{\#inner function code}} \\
\quad\quad\quad\quad\quad\quad  \textbf{\texttt{function transferOwnership(address \_newOwner) public onlyOwner \{}} \\
\quad\quad\quad\quad\quad\quad\quad\quad\quad\quad  \textbf{\texttt{\_transferOwnership(\_newOwner);}} \\
\quad\quad\quad\quad\quad \textbf{\texttt{\}}}
\end{tabular}
} & \\
\hline
        \textbf{Approach} &\textbf{Coment} & \textbf{BLEU-4} &\textbf{METEOR} &\textbf{ROUGE-L}\\  \hline
        Ground Truth &\cellcolor{orange} Transfer ownership of data contract to $\_addr$. $\_addr$ address. &NA &NA &NA \\
        CodeBERT &\makecell[c]{\textcolor{orange}{Transfer ownership} of an \textcolor{orange}{address} to \textcolor{red}{another}. \\ $\_addr$ address \textcolor{orange}{The address} \textcolor{red}{to transfer to}.} &51.20 &28.00 &67.00 \\
        CodeT5 &\makecell[c]{\textcolor{red}{Allows the owner} to \textcolor{orange}{transfer} \textcolor{red}{control of the contract} to an \textcolor{orange}{address}. \\ $\_addr$ \textcolor{red}{The address to transfer ownership to}.} &42.48 &26.64 &47.62 \\
        CodeT5+ &\textcolor{red}{Allows the owner} to \textcolor{orange}{transfer} \textcolor{red}{control} of \textcolor{orange}{data contract} to $\_addr$.  \textcolor{orange}{$\_addr$ The address}. &60.68 &41.67 &60.00 \\
        SCLA &\cellcolor{lightblue} Transfers ownership of the data contract to $\_addr$. &70.77 &72.70 &75.00
        \\\hline
    \end{tabular}}
    \caption{An example illustrating the effectiveness of SCLA.}
    \label{effectiveness_example}
\end{table*}
\textbf{Few-shot Results}. To evaluate the performance of SCLA against SOTA baseline models, we conducted a validation experiment. Since SCLA employs few-shot learning, we tested its performance under Zero-Shot, One-Shot, Three-Shot, and Five-Shot conditions to investigate the number of learning samples required for optimal performance. The results (see Table \ref{sample_learning}) indicate that SCLA initially lags behind the baseline models in Zero-Shot and One-Shot settings. However, starting from Three-Shot, SCLA outperforms the baseline models across all four evaluation metrics: BLEU-4, METEOR, ROUGE-L, and BLEURT. Compared to FMCF, SCLA improved by 17.24\%, 41.26\%, 9.14\%, and 22.00\%, and compared to CodeT5+, the improvements were 21.42\%, 13.55\%, 12.30\%, and 9.21\%. \textbf{Compared to all baseline models, SCLA showed average improvements of 26.7\%, 23.2\%, 16.7\%, and 14.7\% in these metrics}. Performance continued to improve under Five-Shot, although the gains were modest. We also analyzed token consumption to determine the optimal number of few-shot samples (see Table \ref{token_length}). The token consumption for Five-Shot was 48.51\% higher than for Three-Shot, but the average improvement in generated code summarization metrics was only 2.20\%. Therefore, \textbf{Three-Shot provides the best balance between performance and efficiency}.\par
\begin{figure}[h]
    \centering
    \includegraphics[width=0.99\linewidth]{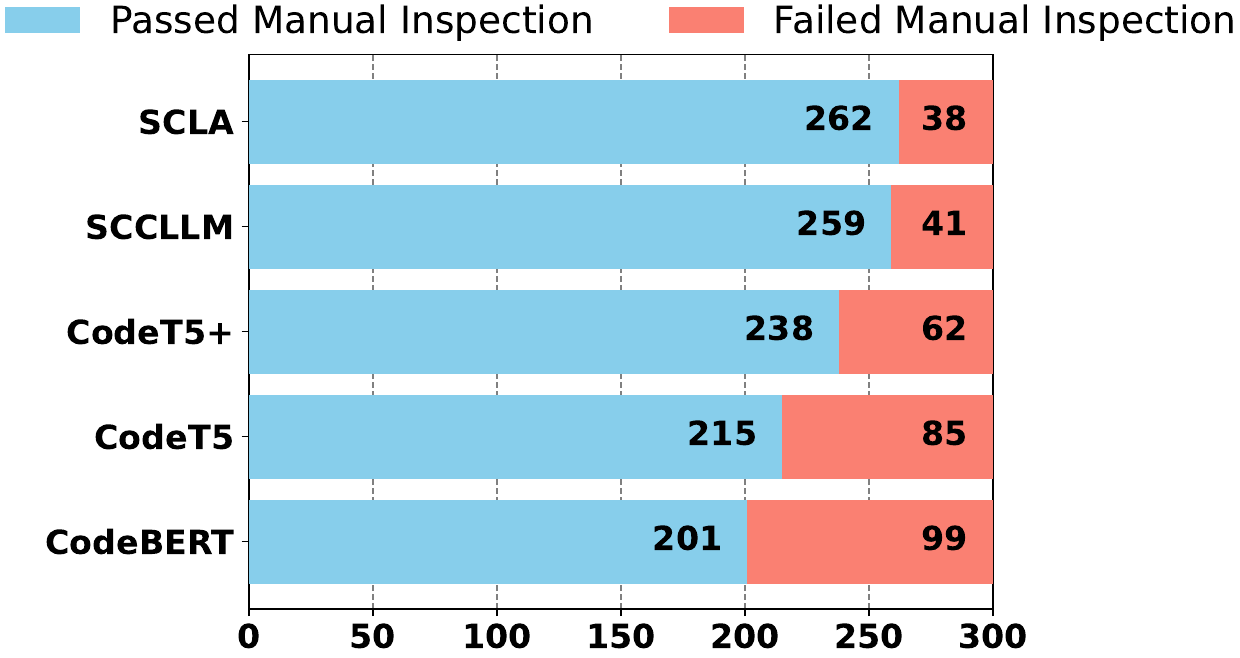}
    \caption{Human Evaluation Results of 300 Code Summarizations Generated by SCLA and the Baseline.}
    \label{Human_Evaluation}
\end{figure}
\subsection{Ablation Study}
We conducted We conducted ablation experiments to assess the contributions of different semantic facts in SCLA. The goal was to quantify the impact of each component on Gemini-1.5-Pro's code summarization performance. Five experiments were performed by selectively removing semantic elements used in semantically enhanced prompts, all under Zero-Shot learning. The results, shown in Table \ref{test}, highlight the importance of inner function ordering, function call graphs, identifiers, and global member variables. Removing inner functions resulted in performance drops of 27.42\%, 1.43\%, 10.93\%, and 4.44\% across key metrics. Excluding the function call graph caused declines of 14.45 (BLEU-4), 5.13 (ROUGE-L), and 1.57 (BLEURT). The removal of identifiers and global member variables led to decreases of 7.72\%, 1.28\%, 1.49\%, and 0.65\%. These findings confirm that inline functions and call graphs are crucial for improving Gemini-1.5-Pro’s code summarization. Additionally, \textbf{experiments underscore the significance of global member variables in maintaining semantic consistency, while function call graphs provide a structural understanding that directly influences summarization accuracy}. Removing these elements disrupts both the coherence and comprehensiveness of the generated summarization. Ultimately, the ablation study validates the necessity of each semantic component for optimizing code summarization performance in scenarios.\par
\subsection{Human Evaluation of Summarization Generated by SCLA and the Baseline}
To assess the summarization generated by SCLA, we randomly selected 300 samples from the smart contract code summarization generated by SCLA and baseline models for manual evaluation. This evaluation focused on similarity, conciseness, and completeness, categorizing the summarization as usable or unusable. To reduce subjectivity and bias, six volunteer evaluators, all Chinese graduate students with experience in smart contract development, were recruited and briefed on the research and evaluation standards. The results, shown in Figure \ref{Human_Evaluation}, reveal that SCLA generated the fewest unusable summarization (38), outperforming all baseline models. \textbf{These findings demonstrate that SCLA is more likely to generate satisfactory smart contract code summarization, reducing the chances of low-quality outputs.}\par
\subsection{Case Study}
Upon reviewing the results, we found that the SCLA prompt includes crucial information for effective summarization. Table \ref{effectiveness_example} highlights the differences between real-world smart contract abstracts and Summarization generated by CodeBERT, CodeT5, CodeT5+, and SCLA. CodeBERT identifies key terms like "transfer," "ownership," and "address," but lacks clarity, with ambiguous pronoun references and repetition of the transfer concept. CodeT5 captures "onlyOwner" but overlooks broader global semantics, rendering the second sentence redundant. CodeT5+ addresses this limitation with more precise terminology, such as identifying the object as a "data contract." \textbf{In contrast, SCLA's Summarization aligns more closely with real-world Summarization, being both more concise and semantically accurate, omitting redundancy for a much clearer, more refined, and contextually precise structure}.\par
\subsection{Generalization Study}
To evaluate the generalization and representativeness of SCLA, we selected 10,000 samples each from Java and Python in the CodeSearchNet dataset~\cite{Codesearchnet}. From these, 1,000 samples per language were randomly chosen as test sets. We compared SCLA's performance against state-of-the-art fine-tuned models—CodeT5, CodeT5+, CodeBERT, and SCCLLM—using BLEU-4, METEOR, and ROUGE-L. On the Java dataset, SCLA outperformed CodeT5+ with improvements of 19\%, 12\%, and 23\% in BLEU-4, METEOR, and ROUGE-L, respectively. On the Python dataset, SCLA showed gains of 7\%, 18\%, and 21\%, respectively. We further compared SCLA to SCCLLM on both datasets using BLEU-4 and BLEURT. With LLMs GPT-4o, Gemini-1.5-Pro, and Claude-3.5-Sonnet, SCLA consistently outperformed SCCLLM, achieving an average improvement of 31.14 in BLEU-4 and 21.38 in BLEURT. \textbf{These results demonstrate that SCLA’s control flow-based prompts generalize effectively to Java and Python, enhancing the LLMs’ ability to capture code structure and improve summarization quality}.\par

\section{Conclusion}
We propose that control flow graphs enhance LLMs' understanding of smart contract code semantics, and experiments confirm their positive impact on code comprehension. Ablation studies assess the contribution of each prompt component to summarization quality. SCLA is a framework that combines LLMs with control flow prompts, outperforming six baseline models. The experiments show that, compared to other baseline models, SCLA significantly improves BLEU-4, METEOR, ROUGE-L, and BLEURT scores with improvements of 26.7\%, 23.2\%, 16.7\%, and 14.7\%, respectively. We also extended SCLA to Java and Python code, further improving summarization and providing new insights for advancing LLM-generated code summarization.
\section*{Limitations}
\begin{figure}[h]
	\centering
   \includegraphics[width=0.9\linewidth,height=3.4cm]{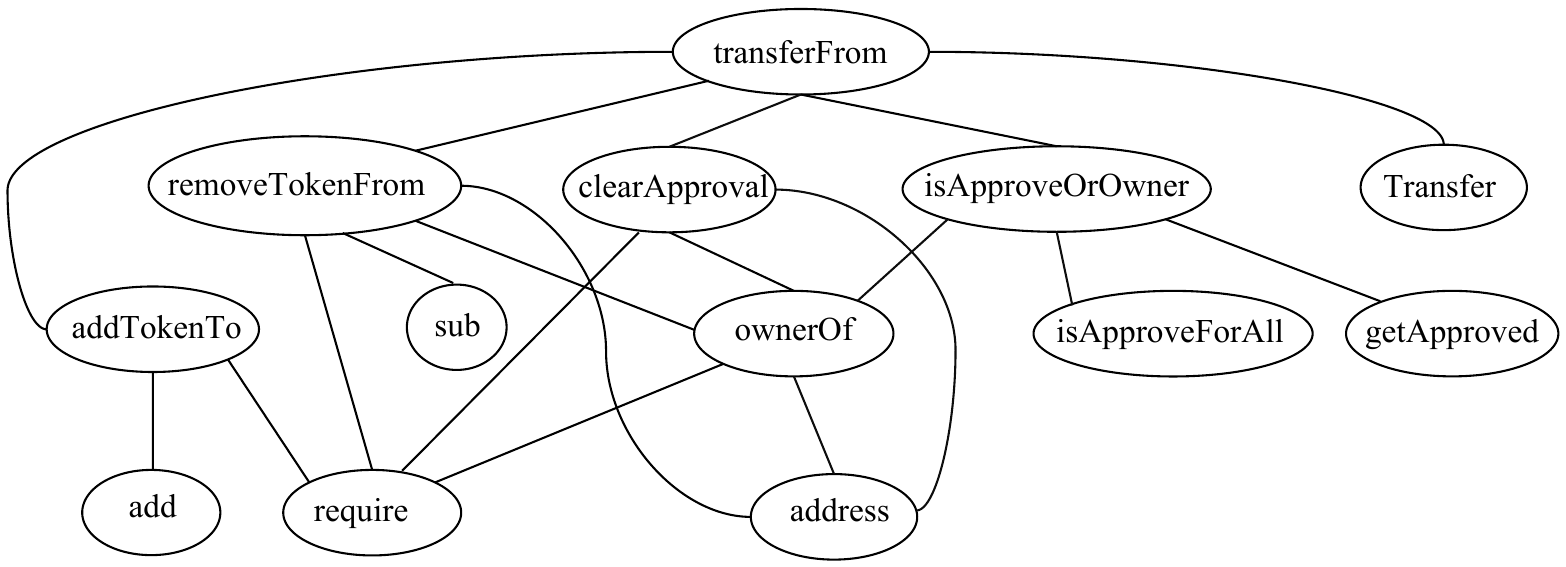}
	\caption{An Example of a Function Call Graph in Which Gemini-1.5-Pro Has Difficulty Understanding the Call Information.}
	\label{transferFrom_example}
\end{figure}

Our framework enhances Gemini-1.5-Pro's understanding using function call graphs. However, Gemini-1.5-Pro struggles with deep call stacks or circular calls. Figure.~\ref{transferFrom_example} shows that circular chains, like transferFrom $\to$ removeTokenFrom $\to$ ownerOf $\to$ isApprovedOrOwner $\to$ transferFrom, confuse the model, leading to misinterpretations and incorrect summarization. In contrast, Gemini-1.5-Pro handles typical tree structures even with a depth of 5. Further research is needed to explore the impact of loop calls and depth on-call interpretation.\par
Another key challenge in using LLMs for smart contract code summarization is the potential exposure of test data during pre-training. Since general-purpose LLMs like GPT-4o and Gemini-1.5-Pro are not publicly accessible, direct verification of this exposure is difficult. Additionally, LLMs' memorization capability can produce artificially high scores if prior summarizations are retained. We also analyzed the effect of few-shot learning on SCLA’s performance in Section~\ref{sample_learning}. Our results show that SCLA outperforms the baseline with a Three-Shot setup, while performance gains plateau at five shots, with a 1.5x increase in computational cost.\par

\bibliography{main}


\end{document}